\documentclass[aps,pra,amsmath,amssymb,floatfix,twocolumn,amsmath,superscriptaddress,twocolumn,nofootinbib,tighten,letterpaper]{revtex4-1}
\usepackage{multirow}
\usepackage{bbold}
\usepackage{subfigure}
\usepackage{color}
\usepackage{mathrsfs}
\usepackage{hyperref}
\usepackage[normalem]{ulem}
\usepackage{bm}

\usepackage{amsfonts, relsize, color}
\usepackage{graphics}
\usepackage{graphicx}
\usepackage{subfigure}
\usepackage{hyperref}
\usepackage{color}
\usepackage{comment}

\renewcommand\vec[1]{\ensuremath\boldsymbol{#1}} 

\begin{document}
\title{Noncrystalline topological superconductors}

\author{Sourav Manna}
\affiliation{Department of Condensed Matter Physics, Weizmann Institute of Science, Rehovot 7610001, Israel}

\author{Sanjib Kumar Das}
\affiliation{Department of Physics, Lehigh University, Bethlehem, Pennsylvania, 18015, USA}

\author{Bitan Roy}
\affiliation{Department of Physics, Lehigh University, Bethlehem, Pennsylvania, 18015, USA}

\begin{abstract}
Topological insulators, featuring bulk-boundary correspondence, have been realized on a large number of noncrystalline materials, among which amorphous network, quasicrystals and fractal lattices are the most prominent ones. By contrast, topological superconductors beyond the realm of quantum crystals are yet to be harnessed, as their nucleation takes place around a well-defined Fermi surface with a Fermi momentum, the existence of which rests on the underlying translational symmetry. Here we identify a family of noncrystalline Dirac materials, devoid of time-reversal (${\mathcal T}$) and translational symmetries, on which a suitable local or on-site pairing yields topological superconductors. We showcase this outcome on all the above mentioned noncrystalline platforms embedded in a two-dimensional flat space. The resulting noncrystalline topological superconductors possess quantized topological invariants (Bott index and local Chern marker) and harbor robust one-dimensional Majorana edge modes, analogs of ${\mathcal T}$-odd $p+ip$ pairing in noncrystalline materials.         
\end{abstract}

\maketitle

\section{Introduction}~\label{sec:inro}

Topological insulators constitute the foundation of topological classification of quantum crystals. In particular, when the electronic wavefunction of a quantum insulator acquires nontrivial geometry in the Brillouin zone, robust gapless modes of charged quasiparticles appear at the interface of topological crystals~\cite{Hasan-Kane-RMP, Qi-Zhang-RMP, ryu:RMP2016, kane-mele-prl2005, qiwuzhang:PRB2006, bhz-science2006, molenkamp-science2007, hsieh-nature2008, fu-prl2011, slager-natphys2013, sato-PRB2014, kruthoff-prx2017, bradlyn-nature2017, tang-nature2019}. Such boundary modes manifest the bulk topological invariant of an electronic wavefunction, a phenomenon named the bulk-boundary correspondence, with quantum Hall insulators standing as its prime examples~\cite{TKNN, Haldane}. Furthermore, topological classification extends beyond the realm of charged fermions, and it is equally applicable to neutral Bogoliubov-de-Gennes (BdG) quasiparticles, opening a rich landscape of topological superconductors~\cite{altland-zirnbauer, ludwig-ryu-furusaki-schnyder, volovik}. The bulk-boundary correspondence is also germane to such paired states, resulting in gapless boundary modes of Majorana fermions.

\begin{figure}[t!]
\includegraphics[width=0.95\linewidth]{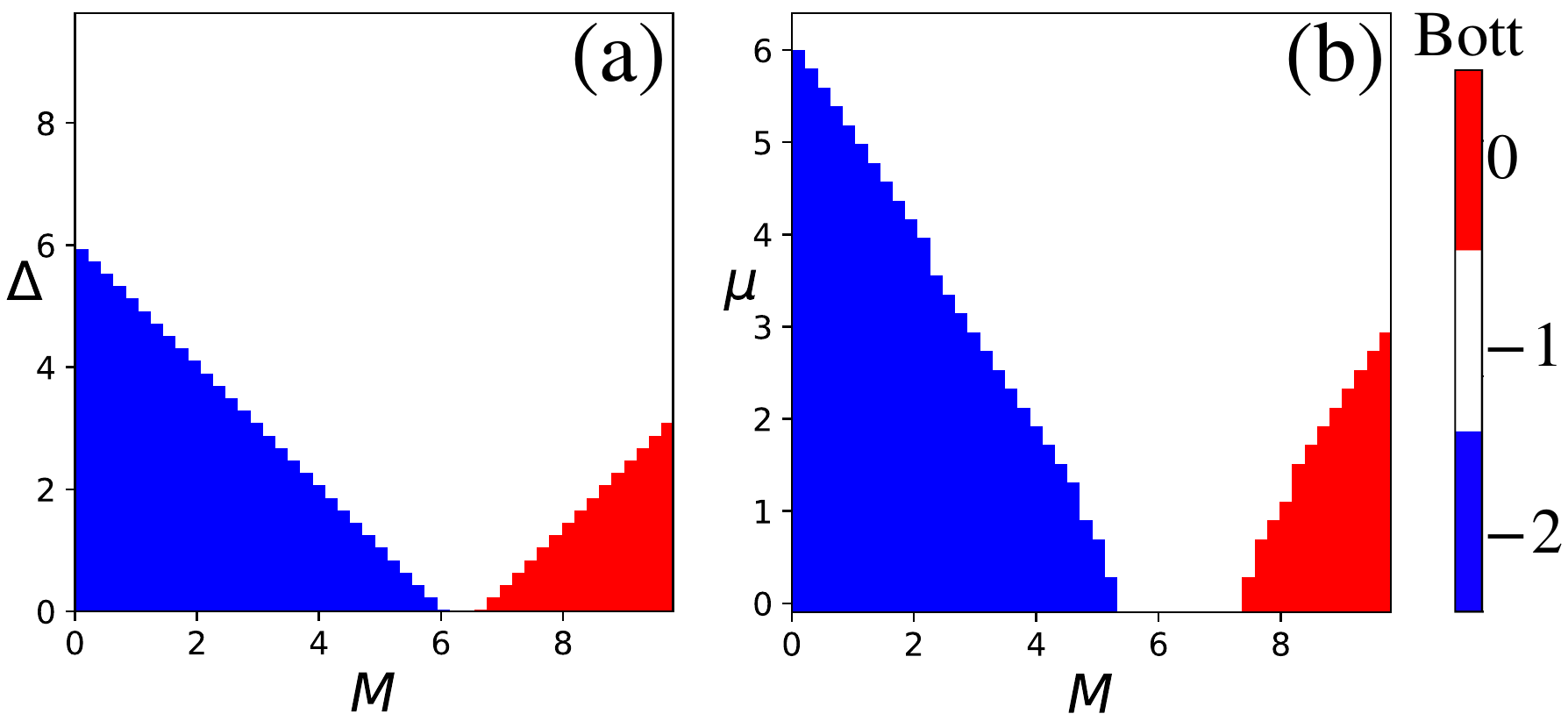}
\caption{Cuts of the global phase diagram of NCTSCs on (a) $(M,\Delta)$ plane for zero chemical doping ($\mu=0$) and (b) ($M, \mu$) plane for a fixed pairing amplitude $\Delta=1$ [Eq.~(\ref{eq:pair})], and for $t=t_0=1$ and $r_0=1$ [Eq.~(\ref{eq:hoppingconfigur})]. We find identical phase diagrams on a square lattice and an amorphous network both containing $400$ lattice sites, Penrose and Ammann-Beenker quasicrystals with 481 and 357 lattice sites, respectively, and third generation Sierpi\'nski carpet and fifth generation glued Sierpi\'nski triangle fractals, containing 512 and 454 lattice sites, respectively, with open boundary condition. This is so because the hopping elements in the normal state are sufficiently long ranged, washing out microscopic details of the underlying noncrystalline order. We find NCTSCs with the topological Bott indices $B=-1$ and $-2$ [Eq.~(\ref{eq:Bott})], both of which support one-dimensional Majorana edge modes [Fig.~\ref{fig:Fig2} and Fig.~\ref{fig:Fig3}], besides the trivial one with $B=0$.            
}~\label{fig:Fig1}
\end{figure}

\begin{figure*}[t!]
\includegraphics[width=1.00\linewidth]{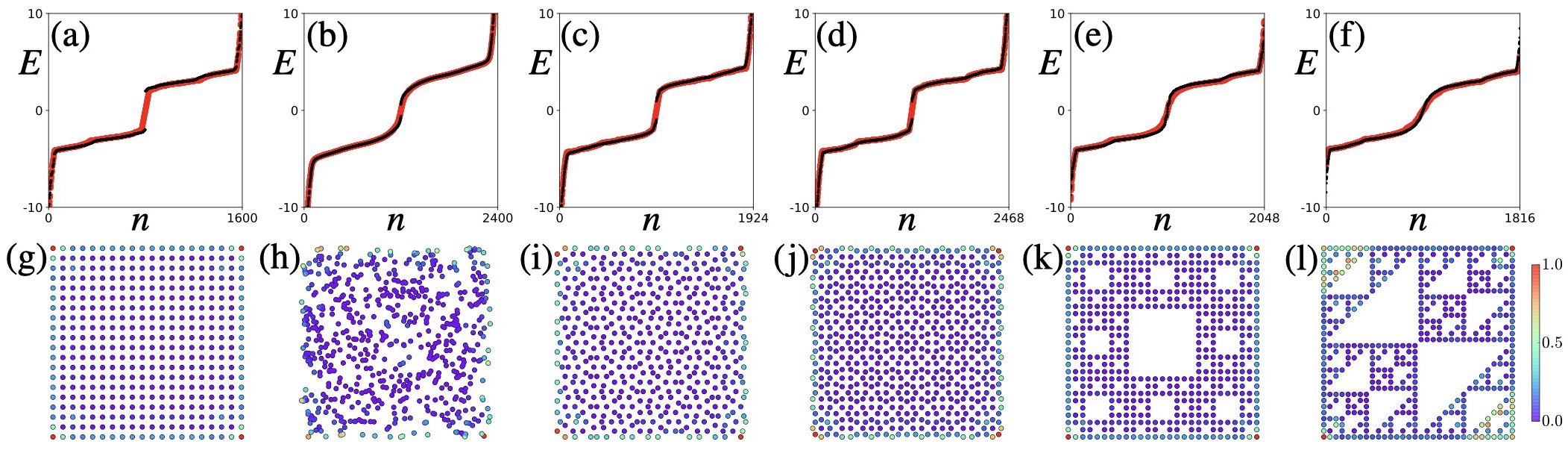}
\caption{Energy spectra of $H^{\rm real}_{\rm pair}$ [Eq.~(\ref{eq:pair})] on (a) a square lattice containing $400$ sites, where $r_0$ is the nearest-neighbor (NN) distance, (b) an amorphous network with 600 sites with $r_0=0.05 \; \times$ linear dimension of the system in each direction, (c) Penrose (481 sites) and (d) Ammann-Beenker (617 sites) quasicrystals, where $r_0$ is the arm length, (e) third generation Sierpi\'nski carpet (512 sites) and (f) fifth generation glued Sierpi\'nski triangle (454 sites) fractal lattices, where $r_0$ is the NN distance, with periodic (black) and open (red) boundary conditions, for $M=2$, $t=t_0=1$, $\Delta=0.5$ and $\mu=0$. We then realize a NCTSC in all these systems with the Bott index $B=-1$ [Fig.~\ref{fig:Fig1}]. Local density of states (LDOS) for two closest to zero energy states (normalized by its maximum value) with open boundary condition on (g) square lattice, (i) Penrose and (j) Ammann-Beenker quasicrystals, (k) Sierpi\'nski carpet and (l) glued Sierpi\'nski triangle fractals, showing sharp edge localization. In (h) we show the total LDOS (normalized by its maximum value) for all the near zero energy states that can only be found with open boundary condition, displaying sharp edge localization. On two fractal lattices the spectra remain qualitatively unchanged with periodic and open geometries due to the inner edges. The LDOS of two closest to zero energy states in open (periodic) fractal lattices shows sharp localization dominantly around the outer (inner) edges. See also Fig.~\ref{fig:Fig3}. We arrive at qualitatively similar results for NCTSCs with $B=-2$ (not shown explicitly). The trivial paired states with $B=0$ is devoid of any near zero energy edge modes.                            
}~\label{fig:Fig2}
\end{figure*}

As over the time it became evident that topological insulators can be realized on noncrystalline materials, such as amorphous networks~\cite{agarwalashenoy, turner:natphys2018, yang:PRL2019, amorphous:costa2019, bailizhang:2020, agarwalaroyjuricic:2020, mukati:2020, grushinamorphous:1, XUPRLamorphous2021a, XUPRLamorphous2021b, grushinamorphous:2, XuSciPost2023}, quasicrystals~\cite{liuQC:PRL2018, liuQC:PRB2018, chenQCHOTI:2020, huaQCHOTI:2020} and fractal lattices~\cite{neupert2018:frac, spaiprem2019:frac, katsnelson2020:frac, souravmanna2020:frac, larsfirtz2020:frac, segev2020:frac, souravmanna2021:frac, mannanandyroy2021:frac, mannajaworowskinielsen2021:fractal, ivakietal2021:fractal, mannaroy2022:ISE}, a question of fundamental and practical importance arises naturally. \emph{Can topological superconductors be realized on noncrystalline materials?} The subtlety of this quest stems from the fact that condensation of electrons into a macroscopic number of Cooper pairs in any real material takes place around a Fermi surface (weak coupling BCS mechanism), defined via a Fermi momentum, which roots in the underlying translational symmetry, as in crystals. Here we provide an affirmative answer to this question, by identifying a family of noncrystalline Dirac materials (NCDMs), encompassing amorphous network, quasicrystals and fractal lattices, on which suitable \emph{local} or \emph{on-site} pairings give rise to nontrivial topology of BdG quasiparticles [Fig.~\ref{fig:Fig1}], and the resulting noncrystalline topological superconductors (NCTSCs) support topological boundary modes of Majorana fermions. See Fig.~\ref{fig:Fig2} and Fig.~\ref{fig:Fig3}.

Although here we exclusively focus on two-dimensional (2D) NCDMs, the underlying mechanism behind the robust realization of NCTSCs is sufficiently general, which can be straightforwardly generalized to three dimensions. Therefore, with the recent experimental realizations of topological insulator in three-dimensional amorphous Bi$_2$Se$_3$~\cite{amorphousbi2se3}, our predicted NCTSCs can be identified in real materials, especially when noncrystalline (topological) insulators are doped, yielding a finite number of charged carriers in the normal state, which at low temperatures can form a Cooper condensate. These findings should promote NCDMs onto a promising platform of Majorana fermion based applications: \emph{Majotronics}.

\subsection{Summary of results}~\label{subsec:summaryintro}

To exemplify these general outcomes, here we consider a 2D time reversal symmetry (${\mathcal T}$) breaking Dirac insulator, constituting the normal state. Depending on the parameter values, such a system can describe either a topological insulator, namely quantum anomalous Hall insulator or a trivial/normal insulator. However, for the discussion of superconductivity and emergent topology of BdG fermions herein, the normal state topology is unimportant. We implement the corresponding model Hamiltonian on a variety of 2D noncrystalline materials besides the square lattice, such as an amorphous network, Penrose and Ammann-Beenker quasicrystals, and Sierpi\'nski carpet and glued Sierpi\'nski triangle fractal lattices, with respective fractal dimensions $d_{\rm frac}=1.89$ and $1.72$~\cite{mannanandyroy2021:frac}. Subsequently, we introduce the only \emph{local} or \emph{on-site} pairing (allowed by the Pauli exclusion principle) in this model and show that it translates into a first-order topological pairing in all the noncrystalline materials, a phase we name as NCTSC. The corresponding global phase diagrams for zero and finite chemical doping are shown in Figs.~\ref{fig:Fig1}(a) and (b), respectively, which besides unfolding NCTSCs also feature trivial pairing.

\begin{figure}[b!]
\includegraphics[width=1.00\linewidth]{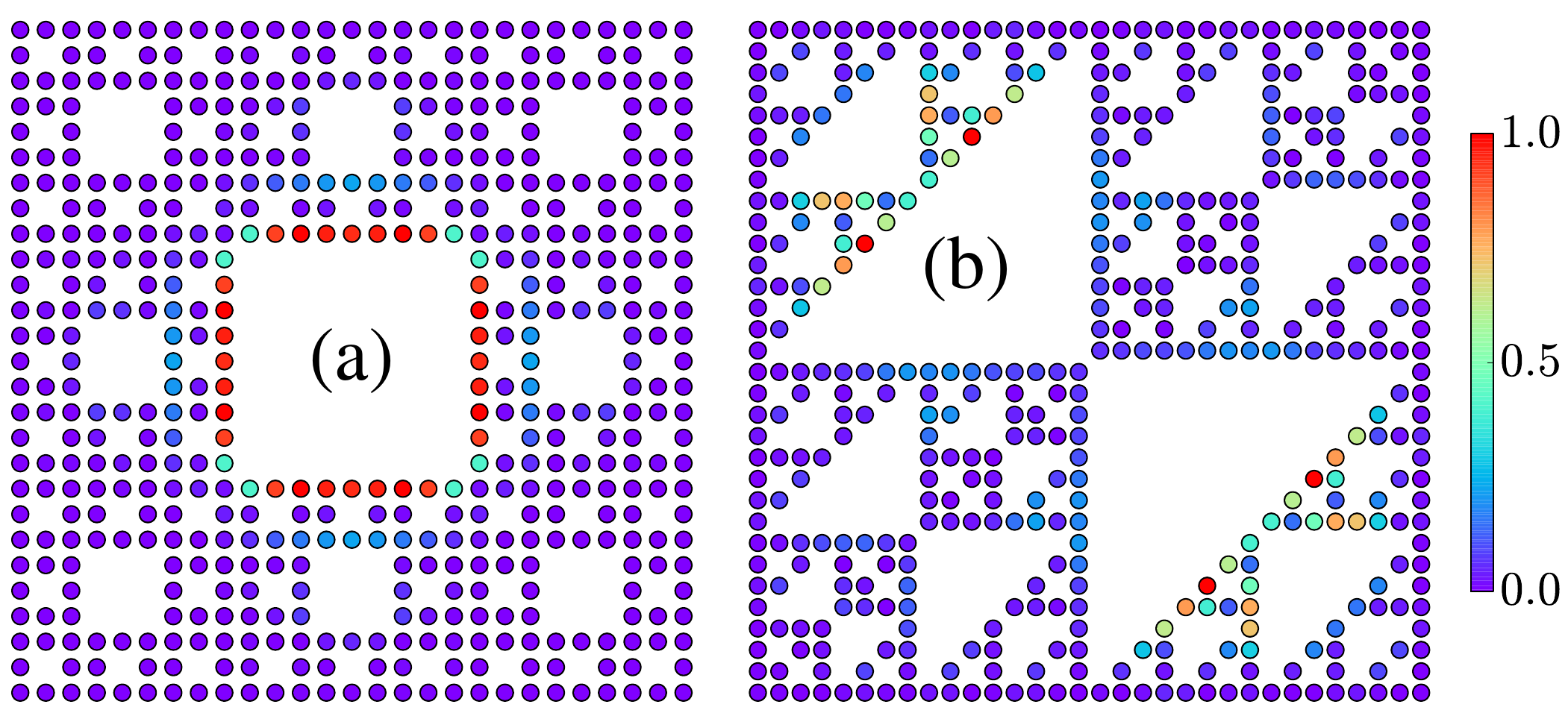}
\caption{The LDOS for two closest to zero energy states on a (a) Sierpi\'nski carpet and (b) glued Sierpi\'nski triangle fractal lattices with periodic boundary condition, displaying their localization dominantly around the inner edges. Compare with Fig.~\ref{fig:Fig2}(k) and (l), respectively.       
}~\label{fig:Fig3}
\end{figure}

\begin{figure*}[t!]
\includegraphics[width=1.00\linewidth]{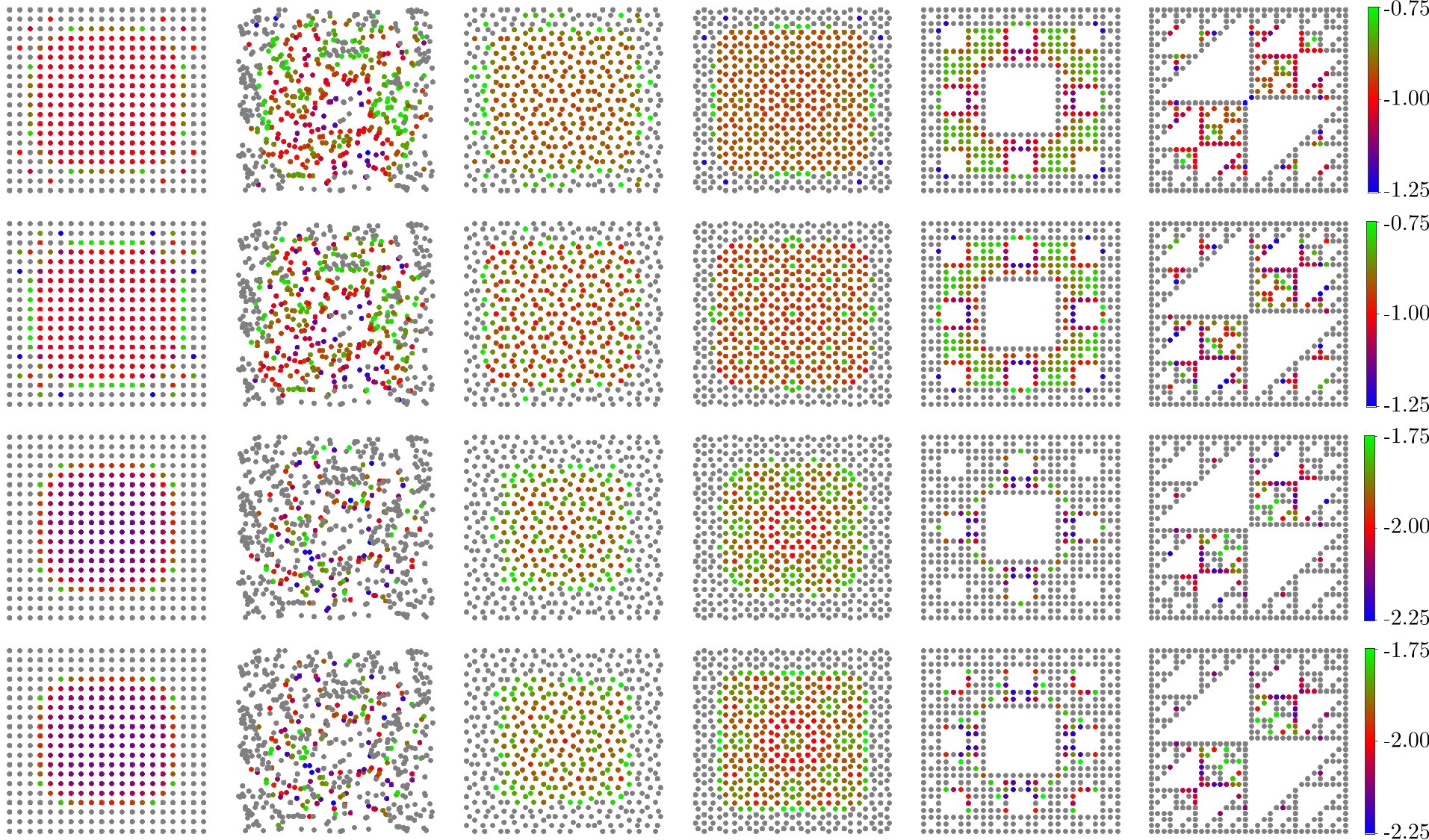}
\caption{Distribution of the local Chern marker $C_{L}(\vec{r})$ [Eq.~\eqref{eq:localchern}] in noncrystalline topological superconducting phases on a square lattice, amorphous system, Pensore and Amman-Beenker quasicrystals, Sierpi\'nski carpet and glued Sierpi\'nski triangle fractals (from left to right) for $\mu=0$, $M=6$ and $\Delta=2$ (top row), $\mu=1$, $M=6$ and $\Delta=1$ (second row), $\mu=0$, $M=2$ and $\Delta=2$ (third row), and $\mu=1$, $M=2$ and $\Delta=1$ (bottom row). For these parameter values the Bott index takes the values $B=-1$ (first two row) and $B=-2$ (last two row) as shown in Fig.~\ref{fig:Fig1}. The system size and the other parameter values are the same as in Fig.~\ref{fig:Fig1}. In each system and row, the values of the local Chern marker on the gray sites fall outside the window specified in the corresponding color bar.             
}~\label{fig:LocalChern}
\end{figure*}

The NCTSCs are characterized by the quantized Bott index ($B$). The global phase diagram supports NCTSCs with $B=-1$ and $-2$, besides a trivial paired state with $B=0$. In addition, topological paired states on all the noncrystalline setups harbor \emph{local} topological regions where the local Chern marker values are peaked around the corresponding quantized Bott index over a certain region in the interior of the system. See Fig.~\ref{fig:LocalChern}. All NCTSCs showcase the bulk-boundary correspondence in terms of one-dimensional (1D) Majorana edge modes in systems with open geometry [Fig.~\ref{fig:Fig2}]. However, exclusively on two fractal lattices with open (periodic) geometry, the edge modes dominantly localize near the outer (inner) edges, as shown in Fig.~\ref{fig:Fig2} (Fig.~\ref{fig:Fig3}). This unique phenomenon on fractal lattices solely stems from its self-similarity symmetry, resulting in inner boundaries in the interior of the system~\cite{fractal:book}.

The edge localization of topological boundary modes is further anchored from the inverse participation ratio (IPR) [Fig.~\ref{fig:Fig4}]. On crystal, amorphous network, and quasicrystals containing $N$ sites, the number of sites at the edges scales as $\sqrt{N}$. Consequently, the IPR for the near zero energy modes scales (almost) linearly with $\sqrt{N}$, ensuring their sharp edge localization. By contrast, due to the inner boundaries there is no such connection between $N$ and the number edge sites on fractals, and the IPR of closest to zero energy modes does not show linear scaling with $\sqrt{N}$ on Sierpi\'nski carpet or glued Sierpi\'nski triangle fractal lattices. This outcome reconciles with $(i)$ energy spectra of NCTSCs on fractal lattices are qualitatively sensitive to boundary conditions [Fig.~\ref{fig:Fig2}(e), (f)] and $(ii)$ on periodic fractal lattices the closest-to-zero energy modes dominantly reside on their inner edges [Fig.~\ref{fig:Fig3}].

\subsection{Organization}~\label{subsec:organization}

The rest of the paper is organized in the following way. In the next section (Sec.~\ref{sec:normalstate}), we introduce the normal state Hamiltonian for the time-reversal symmetry breaking Dirac insulator and implement it on various noncrystalline platforms. In Sec.~\ref{sec:pairing}, we introduce the only local pairing in such a system and construct the effective single-particle BdG Hamiltonian for the paired state. Section~\ref{sec:topoinv} is devoted to the computation of the bulk topological invariants of NCTSCs and the demonstration of the associated bulk-boundary correspondence in terms of Majorana edge modes. Their edge localization in terms of the IPR is discussed in Sec.~\ref{sec:IPR}. Concluding remarks and future directions are presented in Sec.~\ref{sec:summarydiscussion}.

\section{Normal state: Model}~\label{sec:normalstate}

The Hamiltonian for a 2D time-reversal symmetry (${\mathcal T}$) breaking Dirac insulator is~\cite{qiwuzhang:PRB2006, bhz-science2006} 
\begin{align}~\label{eq:DiracIns}
H_{\rm DI}=t \sum_{j=1,2} S_j \tau_j + \bigg[ M-2 t_0 \sum_{j=1,2} \left[ 1- C_j \right] \bigg] \tau_3,  
\end{align}
where $S_j \equiv \sin(k_j a)$, $C_j \equiv \cos(k_j a)$, $\vec{k}=(k_1,k_2)$ is the spatial momenta, $a$ is the lattice spacing, and the vector Pauli matrix ${\boldsymbol \tau}=(\tau_1,\tau_2,\tau_3)$ operates on the orbital space. Using Fourier transformation this model can be implemented on a square lattice. However, when we sacrifice the underlying crystalline order, as in NCDMs, the above model needs to be generalized such that it can be implemented on an arbitrary lattice system. This is accomplished by symmetry analyses of each term appearing in $H_{\rm DI}$ under reflections about the $x$ and $y$ directions, and four-fold rotation in the $xy$ plane. We then replace each momentum dependent factor from $H_{\rm DI}$ by its corresponding term in the real space that transforms identically under all the symmetry operations, leading to  
\begin{eqnarray}\label{eq:normal}
		H^{\rm real}_{\rm DI} &=&\sum_{j \neq k} \frac{F(r_{jk})}{2} \; c^\dagger_j \bigg[  -i t (  C_{jk} \tau_1 + S_{jk} \tau_2) + 2 t_0 \; \tau_3 \bigg] c_k \nonumber \\
		&+& \sum_{j} c_j^\dagger \bigg[(M-4 t_0) \; \tau_3 \bigg] c_j,
\end{eqnarray}
where $C_{jk} \equiv \cos\phi_{jk}$, $S_{jk} \equiv \sin\phi_{jk}$, $c_j=\left[ c_{\alpha j}, c_{\beta} \right]$, and $c_{\tau j}$ is the fermion annihilation operator on site $j$ with orbital $\tau=\alpha$ and $\beta$. Hopping strengths between sites $j$ and $k$, respectively placed at ${\bf r}_j$ and ${\bf r}_k$, are accompanied by  
\begin{equation}~\label{eq:hoppingconfigur}
	F(r_{jk}) = \Theta(r_{jk}-R) \exp \left[1 - \frac{r_{jk}}{r_0} \right],
\end{equation}
ensuring that all the sites are well connected. Here $r_{jk} = |{\bf r}_j - {\bf r}_k|$ is the distance and $\phi_{jk}$ is the azimuthal angle between them, $R$ controls the range of hopping, and $r_0$ is the decay length. For the sake of simplicity, here we typically consider $R$ to be larger than the system size, such that the hopping elements are sufficiently long ranged. This model on any lattice system supports both quantum anomalous Hall and normal insulators, which we do not discuss in any further details.

\begin{figure*}[t!]
\includegraphics[width=0.99\linewidth]{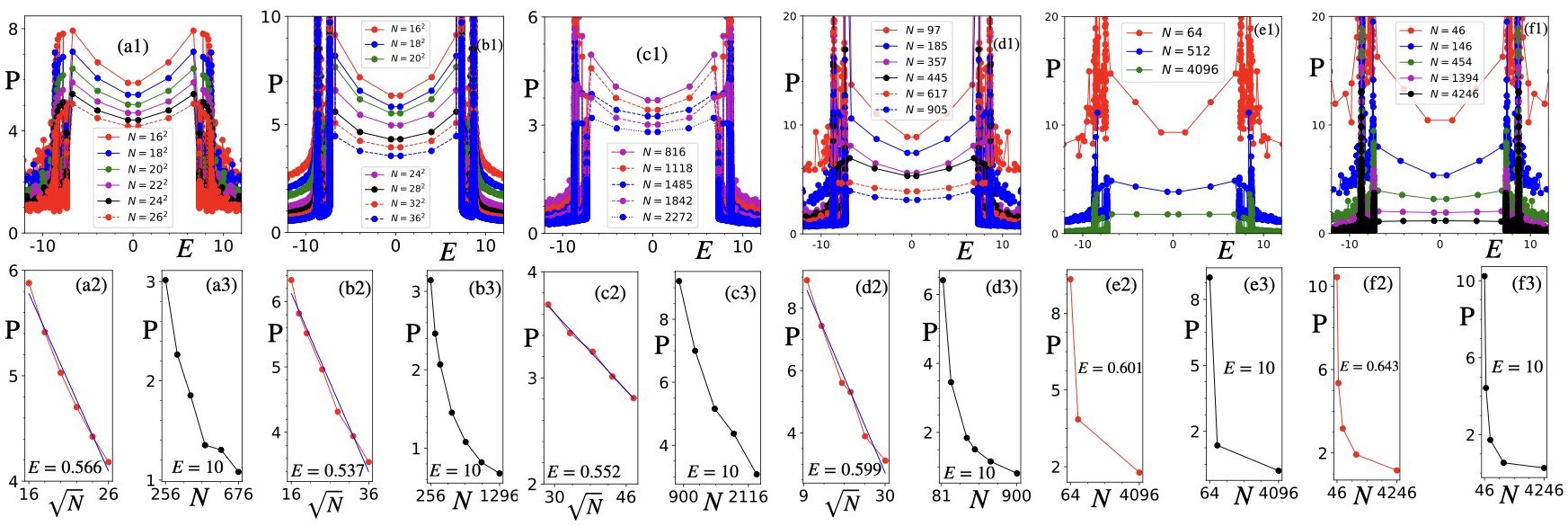}
\caption{Inverse participation ratio (IPR), denoted by $P$ (in units of $10^{-3}$) [Eq.~(\ref{eq:IPR})], on (a) square lattice, (b) amorphous network, (c) Penrose and (d) Ammann-Beenker quasicrystals, (e) Sierpi\'nski carpet and (f) glued Sierpi\'nski triangle fractal lattices, containing $N$ sites (see legends). In each subfigure, panel (1) shows IPR of all the states, panel (2) shows the scaling of the IPR for the closest to zero energy state (edge mode) as a function of $\sqrt{N}$, and panel (3) shows the scaling of a large energy state (bulk mode) as a function of $N$. Except on two fractal lattices, IPR in (2) scales linearly with $\sqrt{N}$, confirming sharp outer edge localization of near zero energy mode [Fig.~\ref{fig:Fig2}]. Absence of such a linear scaling on fractal lattices stems from the existence of the inner boundaries, which host near zero energy modes with periodic geometry [Fig.~\ref{fig:Fig3}].             
}~\label{fig:Fig4}
\end{figure*}

\section{Noncrystalline topological superconductors (NCTSCs)}~\label{sec:pairing}

In principle, $H_{\rm DI}$ can describe a paired state with suitable choice of the spinor basis~\cite{fulgapikulinloring}. For example, the term proportional to $t$ describes a topological $p_x+i p_y$ pairing, when the term proportional to $\tau_3$ yields a Fermi surface, as it happens for $0<M/t_0<8$ on a square lattice. It is, therefore, tempting to conclude that $H^{\rm real}_{\rm DI}$ captures topological pairing on noncrystalline materials. However, such an attempt suffers physical shortcomings. Firstly, $H_{\rm DI}$ does not reveal any microscopic origin of the $p_x+ip_y$ pairing. Even more importantly, the pairing term in $H^{\rm real}_{\rm DI}$ becomes sufficiently long-ranged connecting all the sites with decaying amplitude of the Cooper pairs, which is unphysical as the coherence length of non $s$-wave unconventional superconductors is typically only a few nanometers. Finally, the notion of a Fermi surface in the absence of an underlying translational symmetry becomes moot. To bypass these limitations, we search for a suitable material platform where \emph{on-site} or \emph{local} pairings (immune to structural irregularities) give rise to topological superconductors even in the absence of the underlying translational symmetry.

NCDMs, described by the Hamiltonian $H^{\rm real}_{\rm DI}$, serve this purpose. To accommodate pairings, we Nambu double the theory and introduce the only local or on-site or momentum independent pairing with amplitude $\Delta$. The effective single particle Hamiltonian then reads 
\allowdisplaybreaks[4]
\begin{eqnarray}\label{eq:pair}
		H^{\rm real}_{\rm pair} &=& \sum_{j \neq k} \frac{F(r_{jk})}{2} c^\dagger_j \bigg[  -i t ( C_{jk} \Gamma_{01} + S_{jk} \Gamma_{02}) + 2 t_0 \Gamma_{03}  \bigg] c_k \nonumber \\
		 &+& \sum_{j} c_j^\dagger \bigg[(M - 4 t_0) \Gamma_{03} + \Delta \Gamma_{13} -\mu \Gamma_{30} \bigg]c_j,
\end{eqnarray}
where $\Gamma_{\rho \nu}=\eta_\rho \tau_\nu$, with $\rho,\nu=0,\cdots,3$. The newly introduced Pauli matrices $\{ \eta_\rho \}$ operate on the Nambu or particle-hole index. The chemical potential $\mu$ is measured from the zero energy. The Nambu doubled spinor is $c_j=[c_{\alpha j}, c_{\beta j}, c^\dagger_{\beta j}, c^\dagger_{\alpha j}]$. We implement the above Hamiltonian on a square lattice, as well as on amorphous network, Penrose and Ammann-Beenker quasicrystals, and Sierpi\'nski carpet and glued Sierpi\'nski triangle fractal lattices, prime members of the NCDM family. This construction should be contrasted with recent proposals, where random magnetic impurities were injected on the surface of a pre-existing superconductor to realize NCTSC~\cite{TeemuOjanen2018} and local pairing in a Rashba spin-orbit coupled metal only on Penrose and Ammann-Beenker quasicrystals was introduced to display emergent NCTSC, however only with Bott index $B=1$~\cite{Tohyama2021}.

\section{Topological invariants and bulk-boundary correspondence}~\label{sec:topoinv}

NCTSCs on 2D NCDMs can be identified from the Bott index ($B$). To proceed, we place all the sites of NCDMs within a unit square, and denote their coordinates by $x_i \in [0, 1]$ and $y_i \in [0, 1]$. Then in terms of two diagonal matrices $X_{i,j} = x_i \delta_{i,j}$ and $Y_{i,j} = y_i \delta_{i,j}$, we define two diagonal unitary matrices $U_x = \exp(2\pi i X )$ and $U_y = \exp(2\pi i Y )$. Finally, in terms of the projector (${\mathcal P}$) onto the  filled eigenstates of $H^{\rm real}_{\rm pair}$ up to the Fermi energy ($\mu$), defined as ${\mathcal P}=\sum_{E<\mu} | E \rangle \langle E |$, we compute 
\begin{equation}~\label{eq:Bott}
B=\frac{1}{2 \pi} \text{Im} \left[ \text{Tr} \left[ \ln \left[ V_x V_y V_x^\dagger V_y^\dagger  \right] \right] \right],
\end{equation}
where $V_j = I - {\mathcal P} + {\mathcal P} U_j {\mathcal P}$ for $j=x$ and $y$~\cite{bottindex1}. Two instructive cuts of the global phase diagram of $H^{\rm real}_{\rm pair}$ in the $(M,\Delta)$ plane for $\mu=0$ and the $(M,\mu)$ plane for a fixed $\Delta$, as shown in Fig.~\ref{fig:Fig1}(a) and \ref{fig:Fig1}(b), respectively, unveil NCTSCs with $B=-1$ and $-2$, besides the trivial paired state with $B=0$. Also notice that as the chemical doping is increased, the NCTSCs occupy a larger portion of the phase diagram. Hence, increasing chemical doping enhances the number of carriers in the system and is conducive for the nucleation of NCTSCs, which is thus promising to be observed in doped NCDMs.

The bulk-boundary correspondence for NCTSCs can be established by diagonalizing $H^{\rm real}_{\rm pair}$ on various noncrystalline lattices with periodic and open boundary conditions. Except on two fractal lattices, the energy spectra with these two geometries show clear distinction, especially near the zero energy. Namely, near zero energy modes can only be found for NCTSCs, characterized by $B=-1$ and $-2$, with open geometry. See Fig.~\ref{fig:Fig2} (top row). The LDOS of these modes are highly localized near the outer edges of these systems. See Fig.~\ref{fig:Fig2} (bottom row). By contrast, on fractal lattices the energy spectra are qualitatively insensitive to the boundary conditions and near zero energy modes display strong localization close to the outer (inner) edges of these systems with open (periodic) geometry. See Fig.~\ref{fig:Fig3}. Therefore, NCTSCs like their crystalline counterparts on a square lattice manifest the bulk-boundary correspondence. The fractal lattices in addition display bulk-inner boundary correspondence.

We also complement the Bott index by the local Chern marker in NCTSC phases, displaying local topology, which we compute in the following way. First, we compute the local Chern operator or matrix as~\cite{biancoresta} 
\begin{equation}
\hat{C}_L = - \frac{4 \pi}{A_u} \; \Im \left[ {\mathcal P} \hat{X} {\mathcal Q} \hat{Y} {\mathcal P} \right],
\end{equation} 
where ${\mathcal Q}={\boldsymbol I}-{\mathcal P}$, $\hat{X}$ and $\hat{Y}$ are the position operators in the $x$ and $y$ directions, respectively, and $A_u$ is the area of the unit cell. Subsequently, the local Chern marker at position $\vec{r}$ is given by
\begin{equation}~\label{eq:localchern}
C_L(\vec{r})= \sum^{4}_{j=1} \langle \vec{r}_j | \hat{C}_L | \vec{r}_j \rangle,
\end{equation} 
where $j=1,\cdots, 4$ accounts for two orbitals and Nambu or particle-hole degrees of freedom. The results are shown in Fig.~\ref{fig:LocalChern}. The results are also insensitive to the boundary condition (periodic or open) as two position operators, $\hat{X}$ and $\hat{Y}$ are always nonperiodic. It should be noted that unit cell is well defined only on a square lattice, which is the smallest unit square that is translated to create the whole lattice. Such a notion is absent in all other systems due to the lack of translational symmetry. However, for sufficiently long range hopping in the normal state and with an on-site pairing term, we realize that the global phase diagram is insensitive to the underlying arrangement of the lattice points. See Fig.~\ref{fig:Fig1}. Thus to estimate $A_u$ in any noncrystalline setup (amorphous system, quasicrystals and fractals), we assume that its $N$ sites constitute an effective square lattice for which $A_u$ can readily be obtained. We employ this method to compute the local Chern marker for all NCTSCs.

Note that on a square lattice the local Chern marker is peaked around the corresponding Bott index in the interior of the system, while it deviates substantially from the Bott index value close to the boundaries, as also previously noticed for Chern insulators on crystals~\cite{biancoresta} and quasicrystals~\cite{panigraphi}. The same conclusion is true for amorphous lattice and on quasicrystals. However, on fractal lattice the local Chern marker is concentrated around the corresponding Bott index value, only on the sites that are away from both outer and inner boundaries of the system. This outcome is unique for fractal lattices, featuring inner and outer boundaries due to the self-similarity symmetry. Otherwise, in all these systems the number of sites with local Chern marker around the corresponding Bott index value is almost insensitive to the chemical doping ($\mu$). However, the number of such sites decreases with increasing value of the Bott index. Finally, we note that the distribution of the local Chern marker respects the underlying discrete symmetries of the system, such as four-fold rotational symmetry on square lattice and Sierpi\'nski carpet fractal, five-fold and eight-fold rotational symmetries on Penrose and Ammann-Beenker quasicrystals, respectively, and the inversion symmetry about the 45$^\circ$ diagonal on glued Sierpi\'nski triangle fractal.

\section{Inverse participation ratio (IPR)}~\label{sec:IPR}

To further pin the edge localization of near zero energy modes, we compute the IPR for the generic wavefunction of $H^{\rm real}_{\rm pair}$ with energy $E_j$, given by $| E_j \rangle = \sum_i \psi_{i,j} |i \rangle$ 
\begin{equation}~\label{eq:IPR}
P (j)= \sum_i |\psi_{i,j}|^4,
\end{equation}        
where the index $i$ involves the site, orbital and Nambu indices. The results are shown in Fig.~\ref{fig:Fig4}. Except on two fractal lattices, the IPR for the near zero energy states scales linearly with $\sqrt{N}$ in a system containing $N$ sites. This observation confirms the outer edge localization of these modes as the number of edge sites in all NCDMs scales as $\sqrt{N}$. By contrast, no such scaling of IPR of near zero energy modes can be observed on two fractal lattices due to the existence of the inner edges and the localization of near zero energy modes near them with periodic geometry [Fig.~\ref{fig:Fig3}].

We note that the bulk states residing at the border (along the energy axis) with the delocalized or extended edge modes are most localized with the largest IPR. See Fig.~\ref{fig:Fig4}. This generic observation can be justified qualitatively in the following way. For simplicity, let us assume the confining potential due to finite system size to be a harmonic well and consider only the bulk states therein. The states with lower energies are confined deep within the well, yielding a larger IPR. Whereas states with increasing energy become more delocalized within the bulk, showing a smaller IPR. Naturally, this outcome is insensitive to the exact nature of the confining potential well.

\section{Summary and discussions}~\label{sec:summarydiscussion}

To summarize, here we show that a family of time-reversal symmetry breaking 2D NCDMs, encompassing amorphous network, quasicrystals and fractal lattices, can accommodate 2D NCTSCs that arise from the only allowed (by the Pauli exclusion principle) local or on-site pairing. Consequently, such local NCTSCs are immune to the lack of translational symmetry or structural disorder in NCDMs. They possess a genuine bulk topological invariant, namely the quantized Bott index, encoded by 1D topological edge modes of neutral Majorana fermions, manifesting the bulk-boundary correspondence. A quantized and finite Bott index is also accompanied by nontrivial local Chern markers peaked around the corresponding Bott index value on a finite number of sites buried away from the boundaries (outer and inner) of the system. We also note that chemical doping boosts the propensity of electrons toward the formation of Cooper pairs in the topological channel by increasing the number of carriers in the system. The transition temperature ($T_c$) of amorphous superconductors typically scales differently with the sample thickness in comparison to that of crystals~\cite{TcAmorphous}. Therefore, in a conducive environment, the $T_c$ of at least one NCTSCs can be higher than that of crystalline systems, thereby making Majorana fermion based electronics (Majotronics) operative on noncrystalline materials at higher temperatures. Present work, unambiguously establishing the existence of NCTSCs, motivates a fascinating future exploration of this possibility.

Electronic designer materials~\cite{designer:1, designer:2, designer:3, designer:4}, constitute the most promising platform where our proposed NCTSCs (proximity or phonon mediated) can be observed at low temperatures. In this class of systems, Penrose quasicrystal~\cite{designer:2} and Sierpi\'nski triangle fractal lattice~\cite{designer:4} have been realized, where hopping elements can be tuned at least to a certain degree. Molecular quantum materials are also promising for NCTSCs, where Sierpi\'nski triangle fractal lattices have been engineered~\cite{wu2015:frac}.

The proposed mechanism is sufficiently general, and it can be extended to NCDMs belonging to an arbitrary symmetry class in arbitrary dimensions to demonstrate NCTSCs therein. In particular, our conclusions should be applicable to recently realized three-dimensional topological insulator in amorphous Bi$_2$Se$_3$~\cite{amorphousbi2se3}, especially when it is doped. This specific material is promising as its crystalline counterpart, when doped, possibly harbors odd-parity topological paired states~\cite{TSC:Exp1, TSC:Exp2, TSC:Exp3, TSC:Exp4, TSC:Exp5, TSC:Exp6, TSC:Exp7, TSC:Exp8} that stems from a local or on-site pairing, the exact nature of which, however, still remains a subject of debate~\cite{liangfu:TSC1, liangfu:TSC2, Roy:TSC1, liangfu:TSC3, Roy:TSC2}.

\acknowledgments

S.M.\ thanks the Weizmann Institute of Science, Israel Deans Fellowship through Feinberg Graduate School for financial support. S.K.D.\ was supported by the Startup Grant of B.R.\ from Lehigh University. B.R.\ was supported by NSF CAREER Grant No.\ DMR-2238679. We are thankful to Andr\' as L. Szab\' o and Suvayu Ali for technical support.


\end{document}